\documentclass{PoS}

\usepackage{amssymb, fontenc, times, mathptmx, graphicx}

\newcommand{\isotope}[2]{^{#1}\textrm{#2}}
\newcommand{\reaction}[4]{#1(#2,#3)#4}

\title{AGATA: Gamma-ray tracking in segmented HPGe detectors}

\ShortTitle{AGATA: $\gamma$-ray tracking in segmented HPGe detectors}

\author{\speaker{P.-A. S\"{o}derstr\"{o}m}, A. Al-Adili, J. Nyberg\\
        Department of Physics and Astronomy, Uppsala
  University, SE-75121 Uppsala, Sweden\\
        E-mail: \email{P-A.Soderstrom@physics.uu.se}}
\author{F. Recchia\\
INFN Legnaro, 35020 Legnaro (Padova), Italy}
\author{E. Farnea\\
INFN Padova, 35122 Padova, Italy}
\author{A. Gadea\\
 IFIC CSIC, Valencia, Spain}

\abstract{The next generation of radioactive ion beam facilities, which will give experimental access to many exotic nuclei, are presently being developed. At the same time the next generation of high resolution $\gamma$-ray spectrometers, based on $\gamma$-ray tracking, for studying the structure of these exotic nuclei are being developed. One of the main differences in tracking of $\gamma$ rays versus charged particles is that the $\gamma$ rays do not deposit their energy ``continuously'' in the detector, but in a few discrete steps. Also, in the field of nuclear spectroscopy, the location of the source is mostly well known while the exact interaction position in the detector is the unknown quantity. This makes the challenges of $\gamma$-ray tracking in germanium somewhat different compared to vertexing in silicon detectors. In these proceedings we present the methods for determining the 3D interaction positions in the detector and how these are used to reconstruct the $\gamma$-ray tracks in the AGATA detector array. We also present preliminary simulation results of a proposed in-beam method to measure the interaction position resolution in the germanium detectors.
}

\FullConference{17th International Workshop on Vertex detectors\\
		 July 28 - 1 August  2008\\
		 Ut\"{o} Island, Sweden}

\begin{document}

\section{Introduction}

It is sometimes said that we now are on the brink of the fourth revolution within nuclear spectroscopy \cite{riccirecollection}. The first revolution was the discovery of NaI(Tl) scintillator detectors, by which one could start to measure quite accurately the energy and intensity of $\gamma$-ray transitions in various radioactive nuclei. These measurements later radically improved when one started using semiconductor detectors of germanium instead of NaI(Tl) crystals. The second revolution came with the possibilities to use in-beam $\gamma$-ray spectroscopy together with heavy-ion nuclear reactions. This made it possible to study specific levels of the nuclei of interest. The latest revolution is the development of sophisticated high-resolution spectroscopy arrays of high-purity germanium (HPGe), crowned by EUROBALL \cite{simpseuro} and GAMMASPHERE \cite{gammasphere}. These gave access to very weak signals from high-spin states, unraveling many new nuclear structure phenomena. And now, just around the corner, awaits the second generation facilities for radioactive ion beams (RIB) like NuSTAR/FAIR \cite{npn_nustar,2007JPhG...34..551H} and SIPRAL 2 \cite{2007PrPNP..59...22G} as the fourth revolution. These facilities will make it possible to study very short lived exotic nuclei with extreme values of isospin, located in the \textit{terra incognita} far from the line of $\beta$ stability. Indeed there will be very interesting times for nuclear structure physics in the years to come.

These new exotic nuclei will be produced at the RIB facilities with very low cross sections and in a high $\gamma$-ray background environment, which makes the weak $\gamma$-ray transitions extremely difficult to detect with existing spectrometers. In order to distinguish these rare events, new instruments with higher efficiency and resolving power are required. To meet the new requirements 45 institutes in 12 European countries (as of fall 2008) are collaborating to build the Advanced Gamma Tracking Array (AGATA) \cite{agata_tp,Simpson_npn13-4}. A similar device, the Gamma Ray Energy Tracking Array (GRETA), is also being built in the USA \cite{2003RPPh...66.1095L}.

\section{AGATA\label{sec:agata}}

The AGATA $\gamma$-ray spectrometer will be built in stages and moved between different host laboratories in order to maximally exploit the strengths of the different RIB facilities as they come into operation. A first version of AGATA, the AGATA Demonstrator consisting of 1/12 of the full array, will begin operation at Laboratori Nazionali di Legnaro (LNL) in Italy in the beginning of 2009. The final version of AGATA will consist of a HPGe shell with an inner radius of 23~cm, made of 180 crystals assembled into 60 triple cluster detectors. The total amount of HPGe in AGATA will be 362~kg. Each HPGe crystal has a hexaconical shape and is of the closed-end coaxial type, with the electric field along the radial direction. The crystals are electrically segmented into 36 individual segments, six azimuthal and six axial, plus the central contact (core). The total HPGe solid angle coverage will be 82 \% of $4\pi$. The full-energy peak efficiency of the $4\pi$ array will be $\sim45$~\% for $\gamma$ rays of multiplicity $M_{\gamma} = 1$ and energy 1~MeV, a factor of about four improvement over existing arrays, and $\sim25$~\% for $M_{\gamma} = 30$, which will give several orders of magnitude increase of the resolving power compared to existing arrays \cite{agata_tp}.

This gain in detection power is due to novel techniques to reconstruct the tracks of the $\gamma$ rays that scatter within the array. In order to do this, a three dimensional position resolution of $\lesssim5$~mm of the interaction points (IPs) within the crystal is required. Such a position resolution corresponds to an angular resolution of about 1~degree. This is achieved using digital electronics to sample the pulse shapes and advanced pulse-processing techniques to locate the IPs.

\subsection{Pulse-shape analysis\label{sec:psa}}

\begin{figure}
 \centering
 \includegraphics[width=\textwidth]{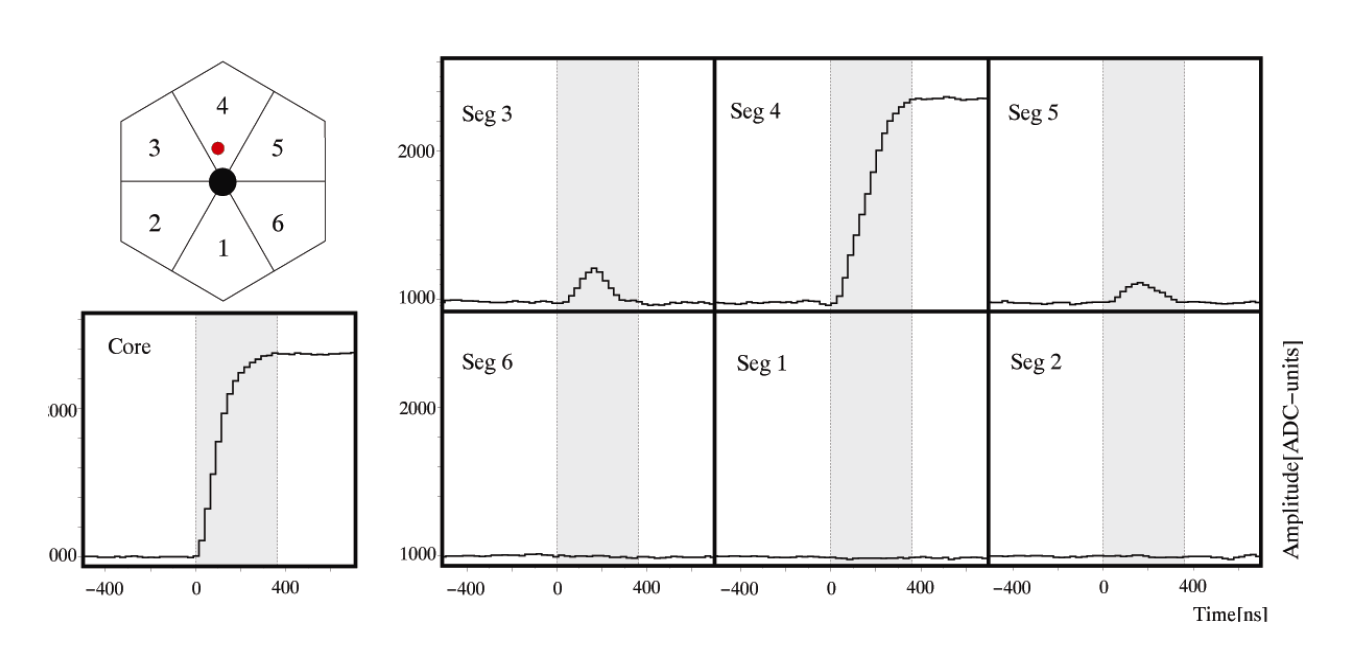}
 \caption{Signals from the core, the segment with the primary hit (Seg 4), and from the mirror charges for a $\gamma$-ray interaction in a six fold segmented HPGe detector. Figure from \cite{agata_tp}.}
 \label{fig:mirrors}
\end{figure}

In order to determine the interaction position, pulse-shape information from the segment which was hit, the mirror charges in the neighboring segments, and the core signal is used, as illustrated in fig.~\ref{fig:mirrors}.
The method to use pulse-shape analysis (PSA) to determine interaction position is, however, not unique for AGATA and GRETA. This is for example done also in MINIBALL, where the azimuthal position is obtained by comparing the amplitudes of the mirror charges \cite{miniball}. The radial position is obtained from the shape of the signal of the hit segment. This is good enough to correct for Doppler effects, but is does not give enough resolution to do $\gamma$-ray tracking. The PSA algorithm that will be implemented in AGATA is the grid search algorithm \cite{lnl_2004_220}, where the sampled pulse shapes are compared to a database of pulse shapes for different interaction positions.

One of the main problem of PSA algorithms is to obtain the database of pulse shapes. When delivered, the crystals are scanned using radioactive sources \cite{Nelson_NIM_A573}. This process gives a very precise database of pulse shapes but it is unfortunately a very slow process. To complement the scanned pulse shapes, codes for calculating them are being developed \cite{simulations_mars}. These calculations are much faster, but the main challenge is to obtain realistic pulse shapes, while dealing with large uncertainties in the impurity concentration and in the modeling of charge mobility \cite{bruyneel}.

\section{Tracking of $\gamma$ rays}

\begin{figure}
 \centering
 \includegraphics[width=0.5\textwidth]{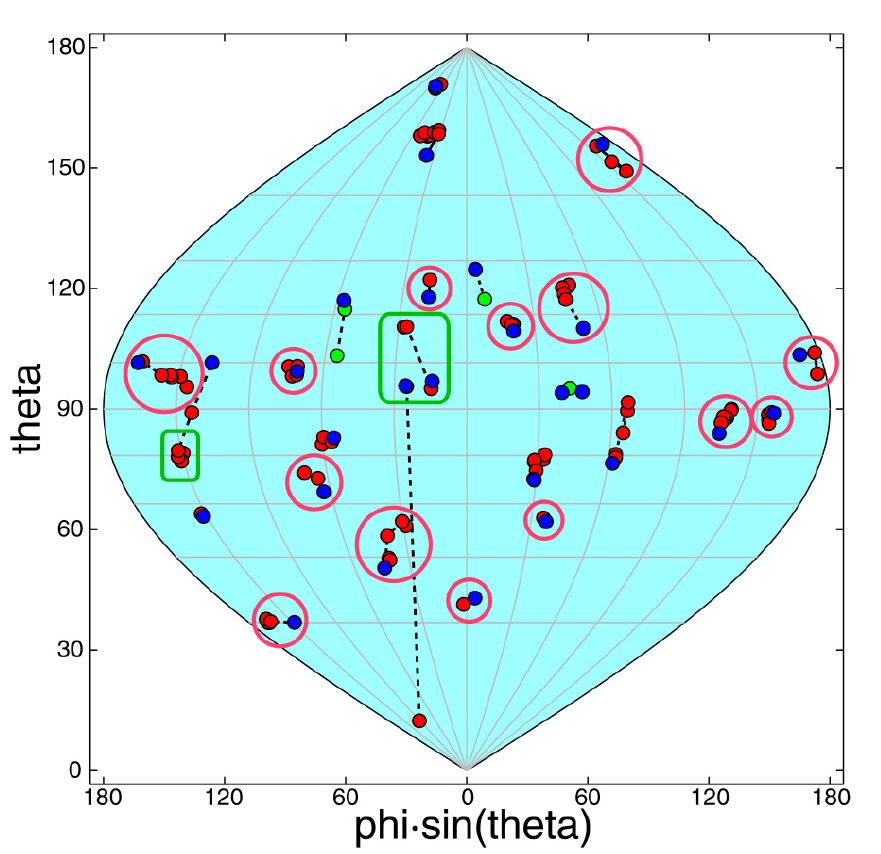}
 \caption{Simulated interaction points of 30 $\gamma$ rays of energy $E_{\gamma} = 1.33$~MeV in the ($\theta$,$\phi \sin \theta$) plane of an ideal germanium shell with an inner radius of 15~cm and an outer radius of 24~cm. Circles are correctly and squares incorrectly identified clusters. Figure from \cite{agata_tp}.}
 \label{fig:worldmap}
\end{figure}

Once the interaction points and the corresponding energies are identified, the different $\gamma$ rays should be disentangled from the ``world map'', see fig.~\ref{fig:worldmap}.
Since the primary interaction mechanism at these energies is a series of Compton scatterings followed by a photoelectric effect, the tracking algorithm is based on the Compton scattering formula
\begin{equation}
 E'_{\gamma} = \frac{E_{\gamma}}{1+\frac{E_{\gamma}}{m_{\mathrm{e}}c^2}(1+\cos \theta)}.\label{eq:compt}
\end{equation}
In order to detect higher and lower energy $\gamma$-rays efficiently the algorithm should also take into account single photoelectric absorption as well as electron-positron pair-production. It is also important that the algorithm can discriminate between $\gamma$ rays and other types of interactions, like neutrons \cite{Ljungvall_NIM_A546,Ljungvall_NIM_A550}. Two main tracking algorithms to be implemented in parallel has been developed and compared carefully \cite{Lopez-Martens_NIM_A533}.

\subsection{Clusterisation or Forward tracking}

One of the algorithms to be used is the clusterisation algorithm, developed by the group at INFN Padova \cite{Bazzacco_NP_A746}. In this algorithm the interactions caused by a certain $\gamma$ ray are assumed to be well localized, which means that the IPs from the same $\gamma$ ray should make up clusters in the $(\theta,\phi \sin \theta)$ plane, see fig.~\ref{fig:worldmap}. After these clusters have been identified, the source location is assumed as the zeroth IP. A first and a second IP is chosen and the $\gamma$-ray energy after scattering is determined from the measured energy depositions in the cluster. This energy is compared to the energy calculated from the Compton scattering formula in eq.~(\ref{eq:compt}) and a figure of merit based on the agreement between the two energies is calculated and if necessary the clusters are rearranged. All possible permutations are evaluated and the one with the best figure of merit is chosen. This procedure is repeated with the identified first IP as a starting point until all IPs in a cluster have been assigned to the track.

\subsection{Backtracking}

The backtracking algorithm was developed by the group at KTH, Stockholm \cite{vanderMarel_NIM_A437,Milechina_NIM_A508}. This algorithm uses the information that the final IP most probable has an energy deposition between 100 to 250 keV. Starting from this assumed final IP, other IPs are searched for within a distance based on the interaction length in germanium for $\gamma$ rays of that energy. This procedure is repeated until the track is terminated by the source location. It is then repeated until no more suitable IPs are available.

\section{In-beam measurements of position resolution}

As mentioned in sec.~\ref{sec:agata}, the position resolution within the crystal is of fundamental importance in order to do $\gamma$-ray tracking. Thus this is an important parameter to measure experimentally, which is planned to be done during the commissioning phase at LNL. One way to do this directly is to use a collimated beam of $\gamma$ rays in order to get a well defined first IP. But this requires a very narrow beam in order not to be limited by the collimation. The consequence of this would be a very time consuming measurement in order to get good statistics. Instead indirect measurements based on imaging techniques \cite{recchia} or in-beam measurements of Doppler correction capabilities are being examined.

\begin{figure}
 \centering
 \includegraphics[angle=90,width=\textwidth]{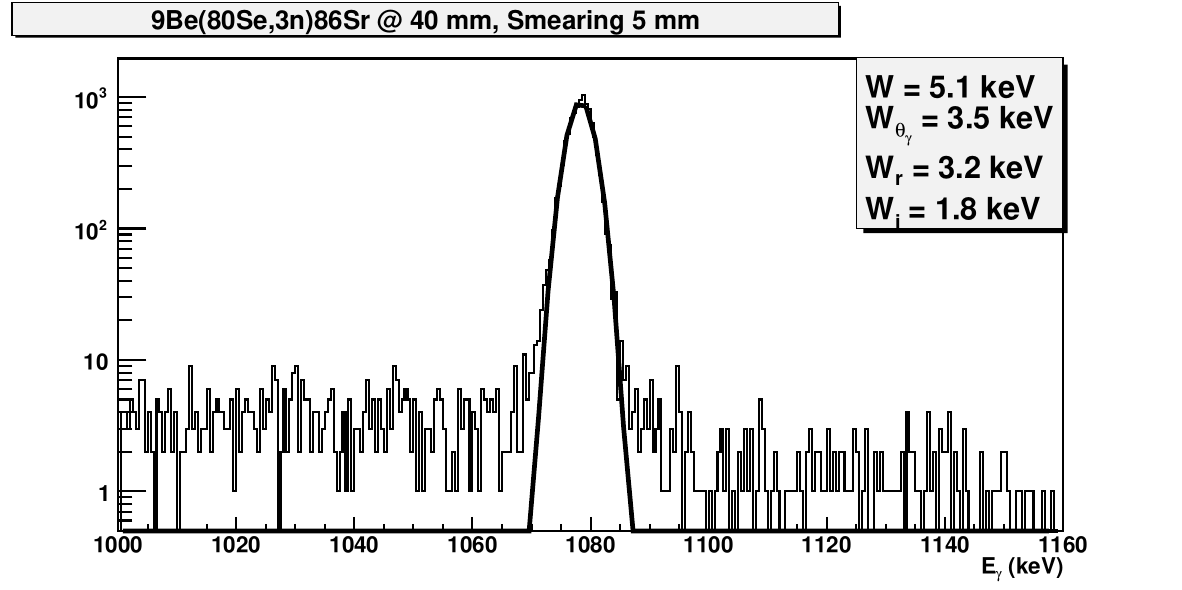}
 \caption{Photo peak from a simulation of the reaction $\reaction{\isotope{9}{Be}}{\isotope{80}{Se}}{3\mathrm{n}}{\isotope{86}{Sr}}$ using the clusterisation algorithm for tracking, a position resolution of 5 mm and a distance between the detector and the source of 40 mm. Here $W_{\mathrm{r}}^2=W_{v_{\mathrm{r}}}^2+W_{\theta_{\mathrm{r}}}^2$.\label{fig:gammapeak}}
 
\end{figure}

\begin{figure}
 \centering
 \includegraphics[width=\textwidth]{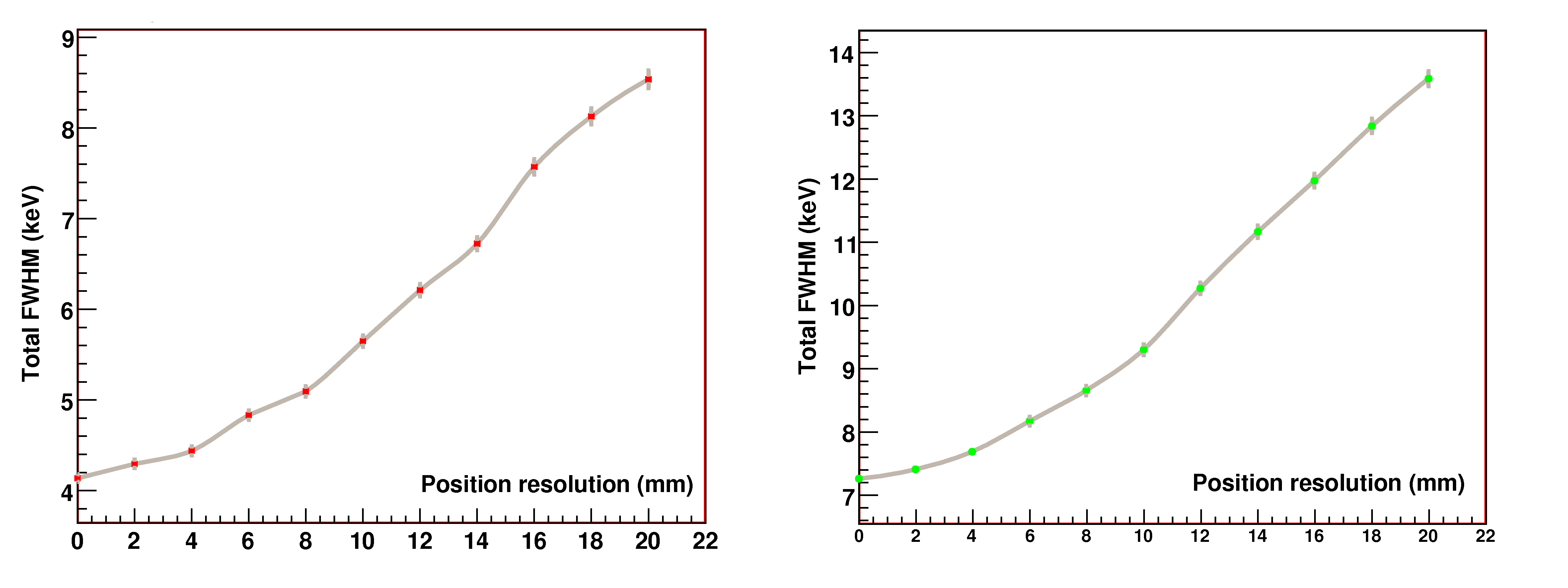}
 \caption{Simulations of the total width, $W$, of the full-energy peak in an AGATA triple cluster detector placed at a distance of 15~cm from the target to the front of the detector and at 90 degrees relative to the beam as a function of the position resolution. In the left panel the reaction $\reaction{\mathrm{d}}{\isotope{51}{V}}{\mathrm{n}}{\isotope{52}{Cr}}$ and in the right panel the reaction $\reaction{\mathrm{d}}{\isotope{37}{Cl}}{\mathrm{n}}{\isotope{38}{Ar}}$ is shown. Figure from \cite{ali}.}
 \label{fig:mcsimus}
\end{figure}

The contribution to the full width at half maximum (FWHM) of the full-energy peak in a $\gamma$-ray spectrum (see fig.~\ref{fig:gammapeak}) can be divided into four parts
\begin{equation}
 W^2 = W_{\mathrm{i}}^2+W_{v_{\mathrm{r}}}^2+W_{\theta_{\mathrm{r}}}^2+W_{\theta_{\gamma}}^2\label{eq:unc},
\end{equation} 
where $W_{\mathrm{i}}$ is the intrinsic resolution of the detector, $W_{v_{\mathrm{r}}}$ and $W_{\theta_{\mathrm{r}}}$ are the contributions from the uncertainties in the velocity and angle of the $\gamma$-ray emitting nucleus (recoil), and $W_{\theta_{\gamma}}$ is the uncertainty in the emission angle of the $\gamma$ ray relative to the beam axis. The last three terms in eq.~(\ref{eq:unc}) are due to effects from the Doppler shift,
\begin{equation}
 E_{\gamma}' = E_{\gamma}\frac{1-\beta\cos\theta}{\sqrt{1-\beta^2}} \approx E_{\gamma}\frac{1-\beta\cos\theta_{\gamma}}{\sqrt{1-\beta^2}}.
\end{equation}
where $\beta = v_{\mathrm{r}}/c$ and $\theta$ is the angle between the recoil and $\gamma$-ray directions. The last expression is valid if the recoil direction is close to the beam direction (0 degrees). Since the $\gamma$-ray emission angle is determined by the position of the first IP in the detector, there is a direct correspondence between $W_{\theta_{\gamma}}$ and the position resolution, $p$, while $W_{\mathrm{i}}$, $W_{v_{\mathrm{r}}}$ and $W_{\theta_{\mathrm{r}}}$ are independent of $p$. 
Since $W_{\theta_{\gamma}}$ depends on the Doppler correction of the $\gamma$-ray energy from $E_{\gamma}$ to $E_{\gamma}'$ as
\begin{equation}
 W_{\theta_{\gamma}}=\frac{\partial E_{\gamma}'}{\partial \theta_{\gamma}} \Delta\theta_{\gamma} = E_{\gamma}\frac{\beta\sin\theta_{\gamma}}{\sqrt{1-\beta^2}}\Delta\theta_{\gamma},
\end{equation} 
with $\Delta\theta_{\gamma}$ being the angular resolution, one would like to have a setup with the detector places as close as possible to the target, as high $\beta$ as possible, and the detector placed at an angle close to 90 degrees. See fig.~\ref{fig:mcsimus} for an example of a simulation, using the AGATA {\sc Geant4} code \cite{agatacode}, showing how $W$ depends on $p$ for two different nuclear reactions.

There is some previous experience using this method to estimate the position resolution of a detector. Kr\"{o}ll et al. have done a similar measurement on the MARS detector \cite{frontiers_ns,mars}, an Italian pre-AGATA detector. The GRETA collaboration has also made measurements of the position resolution of their detectors, using the reaction $\reaction{\isotope{12}{C}}{\isotope{82}{Se}}{4\mathrm{n}}{\isotope{90}{Zr}}$ at  385~MeV. They obtained a $p$ of 4.7~mm \cite{2005NIMPA.553..535D}. Two experiments have also been made with AGATA detectors, one using a single AGATA crystal \cite{eberth} and another using a prototype triple cluster detector \cite{recchia,lnl_2006_231}. Both of these experiments were performed at the tandem accelerator laboratory at University of Cologne.

In the experiment with a single AGATA crystal, the fusion-evaporation reaction $\reaction{\mathrm{d}}{\isotope{37}{Cl}}{\mathrm{n}}{\isotope{38}{Ar}}$ at 70 MeV, and in the experiment with a triple cluster detector, the fusion-evaporation reaction $\reaction{\mathrm{d}}{\isotope{48}{Ti}}{\mathrm{p}}{\isotope{49}{Ti}}$ at 100 MeV was used. The results gave a $p$ of 4.3~mm FWHM at a $\gamma$-ray energy of 2168 keV and a $p$ of 4.4~mm FWHM at a $\gamma$-ray energy of 1382 keV in the two experiments, respectively. The use of protons as the evaporated particle
complicated the setup and the analysis, resulting ultimately
very time consuming, time that will not be available during the
commissioning phase.
Furthermore, the main uncertainty in the estimation process was coming from the modeling of the reaction by the Monte Carlo simulation used to translate the
FWHM of the full-energy
peak to position resolution in the detector.

\begin{figure}
 \centering
 \includegraphics[angle=90,width=\textwidth]{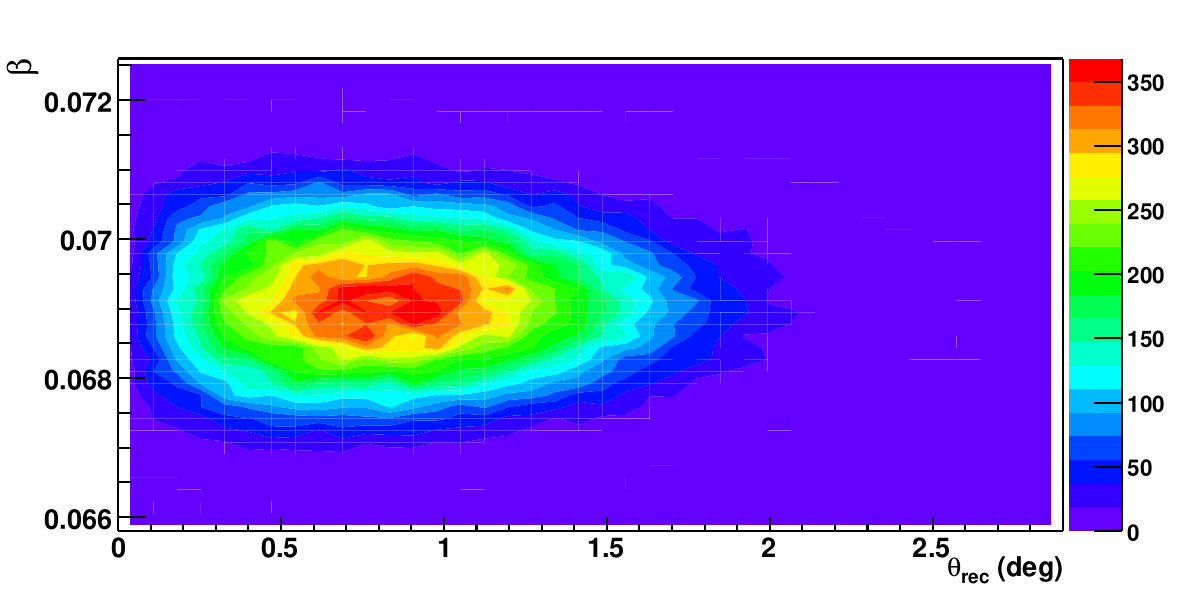}
 \caption{Simulation of the angular and velocity distribution of the recoils in the reaction $\reaction{\isotope{9}{Be}}{\isotope{80}{Se}}{3\mathrm{n}}{\isotope{86}{Sr}}$.\label{fig:angene}}
\end{figure}

The lessons learned from this experiment is that for commissioning, the experimental setup must be simplified even further in order to decrease the analysis time. One also needs to find a Monte Carlo independent way to translate a measured FWHM of the full-energy peak to position resolution. The way to deal with the first issue is to choose another reaction, that evaporates neutrons instead of protons, since these do not have to cross any Coulomb barrier and thus will have lower average energy and the angular deviation of the recoils from 0 degrees with respect to the beam will be as small as possible. A possible reaction with such properties is $\reaction{\isotope{9}{Be}}{\isotope{80}{Se}}{3\mathrm{n}}{\isotope{86}{Sr}}$. Results of a simulation of this reaction are shown in figs.~\ref{fig:gammapeak} and \ref{fig:angene}. As seen, the recoils are well localized in the $(\theta_{\mathrm{r}},\beta)$ plane.

\begin{figure}
 \centering
 \includegraphics[angle=90,width=\textwidth]{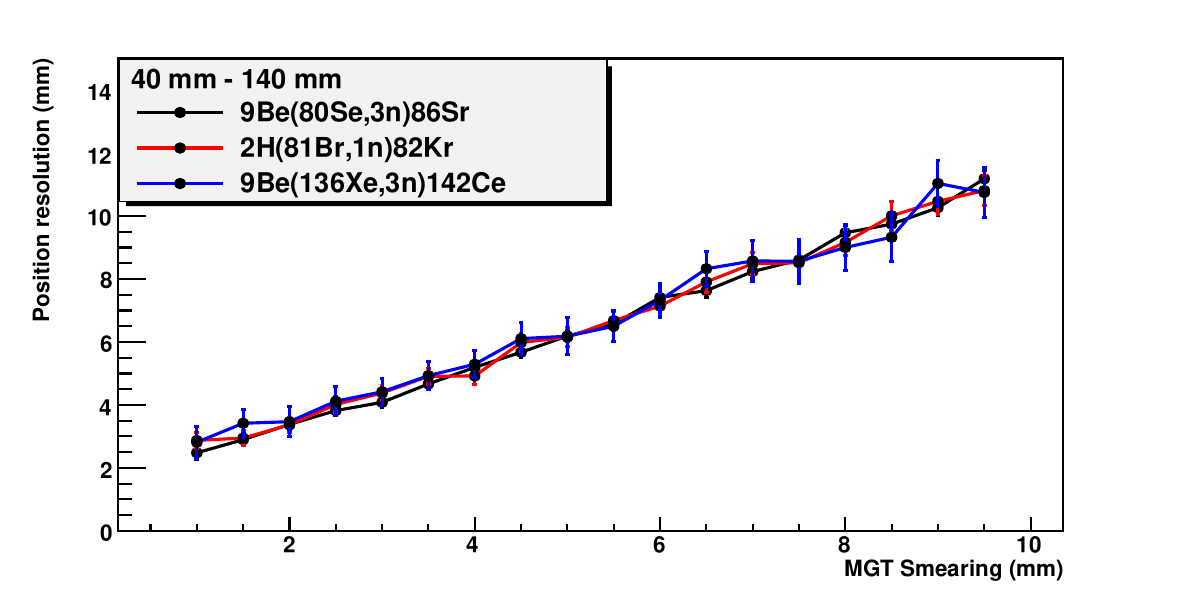}
 \caption{Position resolution as a function of smearing in the tracking program for the reactions $\reaction{\isotope{9}{Be}}{\isotope{80}{Se}}{3\mathrm{n}}{\isotope{86}{Sr}}$, $\reaction{\isotope{2}{H}}{\isotope{81}{Br}}{\mathrm{n}}{\isotope{82}{Kr}}$ and $\reaction{\isotope{9}{Be}}{\isotope{136}{Xe}}{3\mathrm{n}}{\isotope{142}{Ce}}$ for the two distances  $d_{\mathrm{close}}=40$~mm and $d_{\mathrm{far}}=140$~mm.}
 \label{fig:preact}
\end{figure}

A new strategy has been proposed in order to obtain a Monte Carlo independent estimation of the position resolution \cite{recchia}. The idea is that instead of taking a measurement at one distance and compare it to a Monte Carlo simulation, measurements of the total widths $W_{\mathrm{close}}$ and $W_{\mathrm{far}}$ are made at two distances $d_{\mathrm{close}}$ and $d_{\mathrm{far}}$. The position resolution can then be estimated through
\begin{equation}
p^2 = \frac{1}{k^2}\left( W_{\mathrm{close}}-W_{\mathrm{far}}\right) \left( \frac{1}{d_{\mathrm{close}}} -\frac{1}{d_{\mathrm{far}}} \right)^{-1},
\end{equation}
with $k$ being a constant independent of $W$ and $d$. Such a setup has been simulated for different reactions of which preliminary results are shown in fig.~\ref{fig:preact}. The deviation of the estimated values from the simulated ones is due to the fact that AGATA Demonstrator placed close to the target is covering
a wide angle around 90 degrees. Another contribution to this deviation is the uncertainty in the distance to the front face of the detector, due to its curvature radius, when the detector is placed closer to the target than the nominal distance of 235~mm.
Future investigations are foreseen to refine  the estimation procedure. For more details, see ref.~\cite{pa_posi}.

\section{Summary}

In order to study weak transitions in nuclei with extreme values on spin, isospin or temperature, new radioactive ion beam facilities are now being built. These new facilities will be complemented by detector arrays with a great increase in resolving power. The novel technique implemented in these arrays are that of $\gamma$-ray tracking using pulse-shape analysis with digital electronics. The efficiency of the tracking depends, however, critically on the obtainable position resolution of the detector. Some preliminary results on simulations to determine the position resolution of the the AGATA Demonstrator $\gamma$-ray tracking array have been presented.

\section*{Acknowledgments}

We would like to acknowledge the AGATA collaboration for all the effort that has been put into this project. This work was partially supported by the Swedish Research Council and by the  the EURONS project RII3-CT-2004-506065, JRA02: Advanced Gamma Tracking Array (AGATA).

\end{document}